\RequirePackage{amsmath}
\documentclass[runningheads]{llncs}

\usepackage{ecir25-rank-distillm-frame}
\graphicspath{{./ecir25-rank-distillm-figures/}}

\begin{document}

\title{Rank-DistiLLM: Closing the Effectiveness Gap Between Cross-Encoders and LLMs\texorpdfstring{\\}{}for Passage Re-Ranking}
\titlerunning{Rank-DistiLLM}

\author{
  Ferdinand Schlatt\inst{1}\orcidID{0000-0002-6032-909X}
  \and Maik Fr{\"o}be\inst{1}\orcidID{0000-0002-1003-981X}
  \and Harrisen Scells\inst{2,6}\orcidID{0000-0001-9578-7157}
  \and \\
  Shengyao Zhuang\inst{3,4}\orcidID{0000-0002-6711-0955}
  \and Bevan Koopman\inst{3}\orcidID{0000-0001-5577-3391}
  \and Guido Zuccon\inst{4}\orcidID{0000-0003-0271-5563}
  \and \\
  Benno Stein\inst{5}\orcidID{0000-0001-9033-2217}
  \and Martin Potthast\inst{2,6,7}\orcidID{0000-0003-2451-0665}
  \and Matthias Hagen\inst{1}\orcidID{0000-0002-9733-2890}
}
\authorrunning{Schlatt et al.}
\institute{
Friedrich-Schiller-Universit{\"a}t Jena
\and University of Kassel
\and CSIRO \\
\and University of Queensland
\and Bauhaus-Universit{\"a}t Weimar \\
\and hessian.AI
\and ScadDS.AI
% \\\email{ferdinand.schlatt@uni-jena.de}
}

\maketitle
\setcounter{footnote}{0}

\begin{abstract}
Cross-encoders distilled from large language models~(LLMs) are often more effective re-rankers than cross-encoders fine-tuned on manually labeled data. However, distilled models do not match the effectiveness of their teacher~LLMs. We hypothesize that this effectiveness gap is due to the fact that previous work has not applied the best-suited methods for fine-tuning cross-encoders on manually labeled data (e.g., hard-negative sampling, deep sampling, and listwise loss functions). To close this gap, we create a new dataset, Rank-DistiLLM. Cross-encoders trained on Rank-DistiLLM achieve the effectiveness of LLMs while being up to 173~times faster and 24~times more memory efficient. Our code and data is available at {\small\url{https://github.com/webis-de/ECIR-25}}.
\end{abstract}

\section{Introduction}

Cross-encoders~\cite{akkalyoncuyilmaz:2019,nogueira:2020a,xiong:2021} are among the most effective passage re-rankers~\cite{hofstatter:2021,rosa:2022}, but require large amounts of labeled data for fine-tuning. In contrast, large language models~(LLMs) require no further fine-tuning and are often more effective than cross-encoders~\cite{sun:2023,pradeep:2023a,pradeep:2023}. The main drawback of using LLMs is their computational cost, making them impractical for production search engines. However, LLMs can be used to create training data for fine-tuning cross-encoders.

Previous work~\cite{tamber:2023,baldelli:2024} has shown that cross-encoders distilled from LLMs are more effective than cross-encoders fine-tuned on manually labeled data. But the distilled cross-encoders do not reach the effectiveness of their teacher~LLMs because many of the best practices for fine-tuning on manually labeled data were not considered: no ``hard-negative'' sampling was used~\cite{gao:2021,pradeep:2022}, the distillation rankings were not deep enough~\cite{zhuang:2022}, and no listwise losses were used~\cite{gao:2021}.

In this paper, we close the effectiveness gap between distilled cross-encoders and their teacher~LLMs. We analyze the impact of the first-stage retrieval model and the ranking depth on distilled cross-encoders by introducing a new dataset, Rank-DistiLLM, and propose a novel listwise loss function for distillation. As a result, we provide a new distillation method for cross-encoders that is as effective as state-of-the-art ranking LLMs while being orders of magnitude more efficient.

\section{Related Work}
\label{related-work}

MS~MARCO is the most commonly used dataset for fine-tuning cross-encoders as it contains over $500$k~query--passage pairs~\cite{nguyen:2016}. However, most queries only have a single labeled passage. This label sparsity has two implications:
\Ni
the options for suitable loss functions are limited, and
\Nii
``non-relevant'' passages must be sampled heuristically.

Regarding the first implication, listwise losses produce the most effective models~\cite{gao:2021,pradeep:2022,zhuang:2022}. They use a single relevant passage and a set of $k$~heuristically sampled ``non-relevant'' passages. Generally, a higher~$k$ produces more effective models---with~$k=36$ being the highest reported value~\cite{zhuang:2022}. We rely on recent work on memory-efficient fused-attention kernels~\cite{dao:2022,lefaudeux:2022,dao:2023} to fine-tune models on up to $k=100$~passages. Regarding the second implication, ``hard negative'' sampling, i.e., using an effective first-stage retrieval model to sample ``non-relevant'' samples, has produced the most effective models~\cite{gao:2021,pradeep:2022,zhuang:2022}. For instance, models fine-tuned on negatives sampled from ColBERTv2~\cite{santhanam:2022} are more effective than those fine-tuned on negatives sampled from BM25~\cite{robertson:1994}. However, MS~MARCO contains passages that are more relevant than the labeled passage~\cite{arabzadeh:2022}, leading to noisy training data.

To obtain less noisy data, \citet{sun:2023} proposed fine-tuning a cross-encoder on the rankings generated by an LLM applied in zero-shot manner. Models distilled from this dataset are more effective in select scenarios than those fine-tuned on MS~MARCO. More recently, \citet{baldelli:2024} released a dataset that yields more effective cross-encoders by providing more samples per query and using a mixture of first-stage retrieval models. However, an effectiveness gap between a cross-encoder and its teacher~LLMs remains. We investigate if this gap can be closed by applying the insights mentioned above to LLM distillation.

\section{Improving the Fine-tuning of Cross-Encoders}
\label{cross-encoder-fine-tuning}

\paragraph{Preliminaries.}
A cross-encoder outputs contextualized embeddings for every token of a given input sequence $\text{[CLS]} \; q \; \text{[SEP]} \; d \; \text{[SEP]}$, where~$q$ and~$d$ are sequences of query and passage tokens~\cite{devlin:2019}. A linear layer is then applied to the [CLS] token's contextualized embedding to compute the relevance score~$s_{d}$.

\subsection{Traditional Fine-Tuning on MS~MARCO}
\label{sec:traditional-fine-tuning}

\paragraph{Loss}
When fine-tuning cross-encoders on data sampled from MS~MARCO, previous work obtains the most effective models by using the listwise InfoNCE loss~\cite{oord:2019} (also called listwise softmax cross-entropy~\cite{bruch:2019a,zhuang:2022} or localized contrastive estimation~(LCE)~\cite{gao:2021,pradeep:2022}). Given a set of passages~$\mathcal{D}$, of which one is relevant, ${d^+} \in \mathcal{D}$, InfoNCE is defined as:
\[
  \mathcal{L}_{\text{InfoNCE}} = -\log \frac{\exp (s_{d^+})}{\sum_{d \in \mathcal{D}} \exp (s_{d})}.
\]

\paragraph{Data}
To obtain the highest possible effectiveness, $\mathcal{D}$~should be as large as possible. Additionally, the best available first-stage retrieval model should retrieve the other passages $\mathcal{D}^- = \mathcal{D} \setminus \{d^+\}$. Following \citet{pradeep:2022}, we retrieve the top $200$~passages for all MS~MARCO training queries with ColBERTv2~\cite{santhanam:2022} and then randomly sample $7$~hard-negatives.

\subsection{Improved Fine-Tuning on LLM Distillation Data}

\paragraph{Loss}
Instead of a set of passages~$\mathcal{D}$, LLM distillation data consists of a list of passages $\mathcal{R} = (d_1, d_2, \ldots, d_n)$ for a query~$q$ ranked by an LLM. Previous work~\cite{sun:2023,baldelli:2024} uses the pairwise RankNet loss function~\cite{burges:2005} for fine-tuning:
\[
  \mathcal{L}_{\text{RankNet}} = \sum_{i=1}^{n} \sum_{j=i+1}^{n}  \log (1 + \exp(s_{d_j} - s_{d_i})).
\]

To test if listwise loss functions also improve the effectiveness of cross-encoders distilled from LLM rankings, we propose a new loss function based on the Approx family of loss functions~\cite{qin:2010}. Approx loss functions compute a smooth approximation $\hat{\pi}(d)$ of a passage's rank based on all passages' scores. Our new loss, Approx Discounted Rank MSE (ADR-MSE), computes the mean squared error between a passage's actual and approximated rank. Inspired by nDCG, we also apply a logarithmic discount to give higher-ranked passages a higher weight:
\[
  \mathcal{L}_{\text{ADR-MSE}} = \frac{1}{n} \sum_{i=1}^{n} \frac{1}{\log_2(i+1)} (i - \hat{\pi}(d_i))^2.
\]

\paragraph{Data}
\label{sec:llm-distillation-data}
To our knowledge, only two datasets for distilling cross-encoders from LLMs have been released. \citet{sun:2023} released the first dataset (RankGPT) consisting of the top $20$ passages retrieved by BM25~\cite{robertson:1994} and re-ranked by RankGPT-3.5 for $100$k~queries from MS~MARCO. \citet{baldelli:2024} released another dataset (TWOLAR) of the top $30$ passages retrieved by three different retrieval models (BM25, DRAGON~\cite{lin:2023a}, and SPLADE~\cite{formal:2021a}) and re-ranked by RankGPT-3.5 for a total of $20$k-queries from MS~MARCO. TWOLAR produces more effective models, but whether the improved first-stage retrieval models, deeper rankings, or both in combination lead to better effectiveness is unclear.

We create Rank-DistiLLM to systematically investigate the effect of the first-stage retrieval model and the rank depth on a cross-encoder's downstream effectiveness. We retrieve the top 100 passages using BM25 and ColBERTv2 for $10$k randomly sampled queries from the MS~MARCO training set. We then use RankZephyr~\cite{pradeep:2023}, an open-source alternative to RankGPT, to re-rank them. RankZephyr was fine-tuned using the original RankGPT distillation data and additional higher-quality RankGPT-4 distillation data. We refer to this dataset as RankGPT+. To evaluate the effect of ranking depth, we subsample additional datasets by removing all passages that were not within the top $10$, $25$, and $50$ passages of the first-stage retrieval. We release Rank-DistiLLM to the community to facilitate further research.%
\footnote{\url{https://zenodo.org/records/12528410}}

\section{Evaluation}
\label{evaluation}

\paragraph{Experimental Setup}

We mostly follow \citet{pradeep:2022} for fine-tuning cross-encoders. We use HuggingFace~\cite{wolf:2020} ELECTRA\textsubscript{\textrm{BASE}} or ELECTRA\textsubscript{\textrm{LARGE}}~\cite{clark:2020} checkpoints as starting points.%
\footnote{\texttt{google/electra-base-discriminator}}\textsuperscript{,}%
\footnote{\texttt{google/electra-large-discriminator}}
For fine-tuning using MS~MARCO~\cite{nguyen:2016} labels, we data described in Section~\ref{sec:traditional-fine-tuning} and fine-tune for $20$k steps using InfoNCE loss. For fine-tuning on LLM distillation data, we compare our new Rank-DistiLLM datasets with the previously released datasets described in Section~\ref{sec:llm-distillation-data}. We use the TREC Deep Learning 2021 and 2022 tracks~\cite{craswell:2021,craswell:2022} as validation sets and fine-tune until nDCG@10 does not improve for $100$~steps using either RankNet~\cite{burges:2005} or our novel ADR-MSE loss (using $\alpha = 1$). All models are fine-tuned using a batch size of~$32$ and the AdamW~\cite{loshchilov:2019} optimizer with a $10^{-5}$~learning rate. We truncate queries longer than $32$~tokens and passages longer than $256$~tokens. All models are trained on a single NVIDIA A100 40GB GPU. We used the following packages and frameworks to implement models and run experiments: ir\_datasets, ir-measures, Jupyter, Lightning, Lightning~IR, matplotlib, NumPy, pandas, PyTerrier, PyTorch, and SciPy~\cite{kluyver:2016, falcon:2023, schlatt:2024c, hunter:2007, harris:2020, pandas:2024, macdonald:2021, paszke:2019, virtanen:2020, macavaney:2021, macavaney:2022}.

\subsection{In-Domain Effectiveness}

\begin{table}[t]
  \scriptsize
  \centering
  \renewcommand{\tabcolsep}{2.4pt}
  \caption{Comparison of nDCG@10 on TREC DL 2019 and 2020 of baseline models with monoELECTRA directly fine-tuned (Single-Stage) or further fine-tuned from an already fine-tuned model (Two-Stage) on various LLM distillation datasets (RDL: our Rank-DistiLLM dataset). The highest and second-highest scores per task are bold and underlined, respectively. \sig~denotes a statistically significant difference from the underlined monoELECTRA model ($p < 0.05$, Holm-Bonferroni-corrected).}
  \begin{tabular}{@{}llcccc@{\hspace{0.25em}}}
    \toprule
    \bfseries Model                             & \bfseries Dataset        & \multicolumn{2}{c}{\bfseries BM25} & \multicolumn{2}{c}{\bfseries ColBERTv2}                                         \\
    \cmidrule(l@{\tabcolsep}r@{\tabcolsep}){3-4} \cmidrule(l@{\tabcolsep}){5-6}
                                                &                          & DL 19                              & DL 20                                   & DL 19             & DL 20             \\
    \midrule
    First Stage                                 & --                       & 0.480\kernSig                      & 0.494\kernSig                           & 0.732             & 0.724             \\
    \midrule
    RankGPT-4                                   & --                       & 0.713                              & 0.713                                   & 0.766             & 0.793             \\
    RankZephyr                                  & RankGPT+                 & 0.719                              & \underline{0.720}                       & 0.749             & \underline{0.798} \\
    monoT5\textsubscript{\textrm{3B}}           & MS~MARCO                 & 0.705                              & 0.715                                   & 0.745             & 0.757             \\
    RankT5\textsubscript{\textrm{3B}}           & MS~MARCO                 & 0.710                              & 0.711                                   & 0.752             & 0.772             \\
    monoELECTRA\textsubscript{Base}             & MS~MARCO                 & 0.687                              & 0.698                                   & 0.739             & 0.760             \\
    \midrule
    \multicolumn{6}{c}{\em LLM-Distillation -- Single-Stage Fine-Tuning}                                                                                                                          \\
    \midrule
    monoELECTRA\textsubscript{Base}             & RankGPT                  & 0.696                              & 0.666\kernSig                           & 0.690\kernSig     & 0.662\kernSig     \\
    monoELECTRA\textsubscript{Base}             & TWOLAR                   & 0.693                              & 0.669\kernSig                           & 0.754             & 0.730             \\
    monoELECTRA\textsubscript{Base}             & RDL\,\scriptsize{(BM25)} & 0.644\kernSig                      & 0.622\kernSig                           & 0.674\kernSig     & 0.654\kernSig     \\
    monoELECTRA\textsubscript{Base}             & RDL\,\scriptsize{(CBv2)} & 0.709                              & 0.704                                   & \textbf{0.774}    & 0.754             \\
    \midrule
    \multicolumn{6}{c}{\em LLM-Distillation -- Two-Stage Fine-Tuning}                                                                                                                             \\
    \midrule
    monoELECTRA\textsubscript{Base}             & RankGPT                  & 0.664\kernSig                      & 0.634\kernSig                           & 0.477\kernSig     & 0.472\kernSig     \\
    monoELECTRA\textsubscript{Base}             & TWOLAR                   & 0.715                              & 0.706                                   & 0.763             & 0.760             \\
    monoELECTRA\textsubscript{Base}             & RDL\,\scriptsize{(BM25)} & 0.672\kernSig                      & 0.638\kernSig                           & 0.714             & 0.683\kernSig     \\
    \underline{monoELECTRA\textsubscript{Base}} & RDL\,\scriptsize{(CBv2)} & \underline{0.720}                  & 0.711                                   & \underline{0.768} & 0.770             \\
    \arrayrulecolor{lightgray}\midrule\arrayrulecolor{black}
    monoELECTRA\textsubscript{Large}            & RDL\,\scriptsize{(CBv2)} & \textbf{0.733}                     & \textbf{0.727}                          & 0.765             & \textbf{0.799}    \\
    \bottomrule
  \end{tabular}
  \label{tbl:listwise-training}
\end{table}

Table~\ref{tbl:listwise-training} lists nDCG@10 scores of monoELECTRA, a cross-encoder using ELECTRA~\cite{clark:2020} as the backbone encoder, fine-tuned on the data mentioned in Section~\ref{sec:llm-distillation-data}, and evaluated on the TREC DL~2019 and~2020 tasks when re-ranking the top 100~passages retrieved by BM25 and ColBERTv2. We compare our model with \mbox{RankGPT-4}, RankZephyr, monoT5\textsubscript{\textrm{3B}}~\cite{nogueira:2020b}, RankT5\textsubscript{\textrm{3B}}~\cite{zhuang:2022}, and monoELECTRA fine-tuned using MS~MARCO labels. We use a t-test to compute the significance of differences of all models to the best monoELECTRA model fine-tuned using our Rank-DistiLLM dataset with ($p < 0.05$, Holm-Bonferroni-corrected).

\paragraph{Labeled Data vs LLM Distillation}
Our results align with \citet{baldelli:2024} in that a monoELECTRA model only fine-tuned using our ColBERTv2 Rank-DistiLLM dataset is more effective than monoELECTRA fine-tuned using MS~MARCO labels. However, the differences are not statistically significant. Also in line with \citeauthor{baldelli:2024}, we find two-stage fine-tuning, i.e., first fine-tuning on MS~MARCO and continuing to fine-tune on distillation data, to be effective. The two-stage fine-tuned models are more effective than their single-stage fine-tuned counterparts in nearly all cases, but the differences are, again, not statistically significant. In summary, LLM distillation improves effectiveness for in-domain re-ranking, but manual judgements with hard-negative mining still produces competitive models.

The monoT5\textsubscript{\textrm{3B}} and RankT5\textsubscript{\textrm{3B}} models demonstrate that larger models can achieve higher effectiveness when fine-tuned on MS~MARCO labels. To investigate if larger models also improve effectiveness for our Rank-DistiLLM dataset, we fine-tuned a monoELECTRA\textsubscript{Large} model with the two-stage fine-tuning paradigm. The differences to the smaller monoELECTRA\textsubscript{Base} model are again not significant, but the large model does improve effectiveness and is the most effective model, even outperforming RankGPT-4 and RankZephyr, in three out of four cases.

\paragraph{Comparison Between LLM Distillation Datasets}
Models fine-tuned on our ColBERTv2 Rank-DistiLLM dataset are more effective than models fine-tuned on RankGPT and TWOLAR in all cases, irrespective of single-stage or two-stage fine-tuning. Effectiveness improvements are statistically significant compared to RankGPT in all cases and compared to TWOLAR in one case.

We attribute our higher effectiveness to the fact that we use a single high-quality retrieval model for the initial retrieval when generating distillation data. Comparing the model fine-tuned on BM25 Rank-DistiLLM data to the model fine-tuned on ColBERTv2 Rank-DistiLLM data, we find that the latter is substantially more effective. The differences are statistically significant in three of four cases.

\begin{table}[t]
  \scriptsize
  \centering
  \renewcommand{\tabcolsep}{1.26pt}
  \newcolumntype{R}[3]{%
    >{\adjustbox{right=#1,angle=#2,lap=#3-\width}\bgroup}%
    c%
    <{\egroup}%
  }
  \newcommand{\rot}[4]{\multicolumn{1}{R{#1}{#2}{#3}}{#4}}
  \caption[]{Effectiveness in nDCG@10 micro-averaged across all queries from a collection from the TIREx framework~\cite{frobe:2023}. The macro-averaged geometric mean is computed across all corpora. The highest and second-highest scores per corpus are bold and underlined, respectively. \sig~denotes a statistically significant difference from the underlined monoELECTRA model ($p < 0.05$, Holm-Bonferroni-corrected). The dataset used to fine-tune each model is provided for context (RDL: our Rank-DistiLLM dataset).\protect\footnotemark
  }
  \begin{tabular}{@{}llcccccccccccccc@{}}
    \toprule
    \raisebox{-7ex}{\bfseries Model}
                                          & \raisebox{-7ex}{\bfseries Dataset}
                                          & \rot{5.2em}{-30}{1.5em}{\bfseries Antique}
                                          & \rot{5.2em}{-30}{1.5em}{\bfseries Args.me}
                                          & \rot{5.2em}{-30}{1.5em}{\bfseries ClueWeb09}
                                          & \rot{5.2em}{-30}{1.5em}{\bfseries ClueWeb12}
                                          & \rot{5.2em}{-30}{1.5em}{\bfseries CORD-19}
                                          & \rot{5.2em}{-30}{1.5em}{\bfseries Cranfield}
                                          & \rot{5.2em}{-30}{1.5em}{\bfseries Disks4+5}
                                          & \rot{5.2em}{-30}{1.5em}{\bfseries GOV}
                                          & \rot{5.2em}{-30}{1.5em}{\bfseries GOV2}
                                          & \rot{5.2em}{-30}{1.5em}{\bfseries MEDLINE}
                                          & \rot{5.2em}{-30}{1.5em}{\bfseries NFCorpus}
                                          & \rot{5.2em}{-30}{1.5em}{\bfseries Vaswani}
                                          & \rot{5.2em}{-30}{1.5em}{\bfseries WaPo}
                                          & \rot{5.2em}{-30}{1.5em}{\bfseries G.\ Mean}                                                                                                                                                                                                                                                                                                                                           \\
    \midrule
    First Stage                           & --                                           & .516\kernSig     & .405                     & .177             & \textbf{.364}\kernSig    & .586\kernSig          & \textbf{.012}    & .424\kernSig             & .259\kernSig             & .467\kernSig          & .385                  & .281\kernSig             & .447\kernSig     & .364\kernSig             & .286             \\
    \midrule
    RankZephyr                            & RGPT+                                        & .534\kernSig     & .364                     & .213             & .303                     & \textbf{.767}\kernSig & .009             & \underline{.542}\kernSig & \underline{.349}\kernSig & \underline{.560}      & \textbf{.460}\kernSig & .314                     & .512             & \textbf{.508}\kernSig    & \underline{.320} \\
    monoT5\textsubscript{\textrm{3B}}     & MSM                                          & .590             & \underline{.415}\kernSig & .188             & .323\kernSig             & .649                  & \underline{.011} & .526\kernSig             & .345\kernSig             & .529                  & .395                  & \underline{.319}\kernSig & .474\kernSig     & .469                     & .313             \\
    RankT5\textsubscript{\textrm{3B}}     & MSM                                          & \textbf{.598}    & \textbf{.421}\kernSig    & \textbf{.227}    & \underline{.336}\kernSig & .713                  & .010             & .538\kernSig             & \textbf{.353}\kernSig    & .528                  & .406                  & \textbf{.323}\kernSig    & .459\kernSig     & .468                     & \textbf{.322}    \\
    \midrule
    mELEC\textsubscript{Base}             & MSM                                          & .517\kernSig     & .326\kernSig             & .164\kernSig     & .252\kernSig             & .667                  & .008             & .436\kernSig             & .254\kernSig             & .491\kernSig          & .366\kernSig          & .268\kernSig             & .456\kernSig     & .406                     & .267             \\
    mELEC\textsubscript{Base}             & TWOLAR                                       & .576\kernSig     & .305\kernSig             & .186             & .292                     & .653\kernSig          & .009             & .486\kernSig             & .263\kernSig             & .523                  & .406                  & .281\kernSig             & .519             & .434                     & .288             \\
    \underline{mELEC\textsubscript{Base}} & RDL\,\scalebox{.8}{(CBv2)}                   & \underline{.593} & .375                     & .209             & .295                     & .692                  & .010             & .507                     & .305                     & .541                  & .399                  & .306                     & \underline{.522} & .458                     & .309             \\
    mELEC\textsubscript{Large}            & RDL\,\scalebox{.8}{(CBv2)}                   & .575\kernSig     & .369                     & \underline{.221} & .313\kernSig             & \underline{.716}      & .008             & \textbf{.546}\kernSig    & .344\kernSig             & \textbf{.572}\kernSig & \underline{.419}      & .316\kernSig             & \textbf{.526}    & \underline{.504}\kernSig & .318             \\
    \bottomrule
  \end{tabular}
  \label{tbl:tirex-results}
\end{table}

\subsection{Out-of-Domain Effectiveness}
Table~\ref{tbl:tirex-results} lists the effectiveness on all corpora from the TIREx framework~\cite{frobe:2023,hashemi:2020,clarke:2009,clarke:2010,clarke:2011,clarke:2012,collins-thompson:2013,collins-thompson:2014,bondarenko:2021,bondarenko:2022,voorhees:2020,wang:2020,cleverdon:1991,voorhees:1998,voorhees:1999,voorhees:2004,craswell:2002,craswell:2003,craswell:2004,clarke:2004,clarke:2005,buttcher:2006,hersh:2004,hersh:2005,roberts:2017,roberts:2018,craswell:2019,craswell:2020,boteva:2016}. It shows that the monoELECTRA\textsubscript{Large} model two-stage fine-tuned using our ColBERTv2 Rank-DistiLLM dataset is only marginally less effective than the state-of-the-art RankZephyr and largest RankT5 models. Comparing the TWOLAR dataset to the Rank-DistiLLM dataset shows that our dataset produces a more effective model on 12~of~13 corpora and the differences are significant on $6$~corpora. Our Rank-DistiLLM dataset therefore closes the effectiveness gab between distilled cross-encoders and their teacher LLMs for both in-domain and out-of-domain re-ranking.

\subsection{Data Ablation}
\begin{figure}[t]
\begin{minipage}[t]{0.48\textwidth}
  \vspace{0pt}
  \centering
  \captionof{figure}{Average effectiveness on TREC DL 2019 and 2020 for models fine-tuned on subsamples of RankDistiLLM using different depths and numbers of samples.}
  \includegraphics[width=\linewidth]{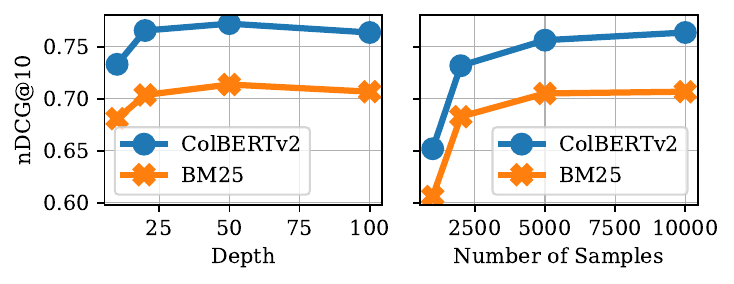}
  \label{fig:data-ablation}
\end{minipage}
\hfill
\begin{minipage}[t]{0.48\textwidth}
  \vspace{0pt}
  % \begin{table}
\scriptsize
\centering
\captionof{table}{Time in seconds and memory consumption in gigabytes for re-ranking 100~passages.}
% \caption{Time in seconds and memory consumption in gigabytes for re-ranking 100~passages from TREC DL 2019 and 2020 for different models.}
\setlength{\tabcolsep}{5.5pt}
\begin{tabular}{@{}lrrr@{}}
  \toprule
  \bfseries Model             & \bfseries Param. & \bfseries Time & \bfseries Memory \\
  \midrule
  RankGPT-4                   & N/A              & 20.234         & N/A              \\
  RankZephyr                  & 7B               & 24.047         & 15.48            \\
  monoT5\textsubscript{3B}    & 3B               & 0.998          & 29.36            \\
  RankT5\textsubscript{3B}    & 3B               & 0.942          & 29.04            \\
  \midrule
  m.ELEC\textsubscript{Base}  & 110M             & 0.139          & 1.18             \\
  m.ELEC\textsubscript{Large} & 330M             & 0.215          & 2.69             \\
  \bottomrule
\end{tabular}
\label{tab:efficiency}
% \end{table}
\end{minipage}
\end{figure}

Figure~\ref{fig:data-ablation} shows that effectiveness peaks at $50$~samples per query and slightly decreases at $100$~samples. We attribute the lower effectiveness at $100$~samples to the training data containing fewer relevant documents at lower depths but further work is necessary to fully investigate this effect. When downsampling the number of training samples, we achieve the highest effectiveness using all $10$k queries. Since monoELECTRA\textsubscript{Large} can reach the effectiveness of RankZephyr in two-stage fine-tuning, we assume $10$k queries are sufficient in this case. However, more data may improve effectiveness in single-stage fine-tuning.

\footnotetext{This table very slightly differs from the table in the \href{https://link.springer.com/chapter/10.1007/978-3-031-88714-7_31}{official proceedings}: it correctly reports micro-averaged results and the geometric mean.}

\subsection{Listwise Fine-Tuning}

Our newly proposed listwise ADR-MSE loss function produces a marginally less effective model at 0.002 (single-stage) and 0.001 (two-stage) lower nDCG@10 compared to using RankNet averaged across TREC DL 2019~/~2020 and BM25~/~ColBERTv2 for initial retrieval. Since the effectiveness is practically equal, we conclude that listwise loss functions are (currently) unnecessary for distillation from LLMs. Future more complex tasks may benefit from listwise distillation.

\subsection{Efficiency}

Table~\ref{tab:efficiency} reports the model size in parameters, latency in seconds, and GPU memory consumption in gigabytes. Our monoELECTRA models use vastly smaller backbone encoder models compared to monoT5\textsubscript{\textrm{3B}} and RankT5\textsubscript{\textrm{3B}}, reducing latency and memory consumption. Our monoELECTRA\textsubscript{Large} model is around $4$~times faster and needs around 10\% of the memory compared to monoT5\textsubscript{\textrm{3B}} and RankT5\textsubscript{\textrm{3B}}. While RankZephyr uses an even larger backbone encoder model, the required memory is lower than for monoT5\textsubscript{\textrm{3B}} and RankT5\textsubscript{\textrm{3B}} because it only re-ranks $20$~passages at a time. The model still requires around $5.7$ times the amount of memory as monoELECTRA\textsubscript{Large}. Furthermore, the windowed ranking strategy necessitates that some passages are scored multiple times, leading to very poor latency. RankZephyr is around $110$ times slower than monoELECTRA\textsubscript{Large} at comparable effectiveness. This poor latency also applies to RankGPT-4 since it uses the same ranking strategy.

\section{Conclusion}

Using our new Rank-DistiLLM datset, we have systematically investigated several aspects of distilling cross-encoders from LLM~rankings. Our findings indicate that rankings of the top-50~passages for 10,000~queries suffice to achieve competitive effectiveness compared to LLMs, but the passages need to be sampled using a very effective first-stage retrieval model. By first fine-tuning on MS~MARCO labels and then further on Rank-DistiLLM, our best model is more effective than previous cross-encoders and matches the effectiveness of LLMs for in- and out-of-domain re-ranking while being orders of magnitude more efficient.

{\fontsize{9pt}{11pt}\selectfont 
\subsubsection*{Acknowledgements}
This publication has received funding from the European Union's Horizon Europe research and innovation programme under grant agreement No 101070014 (OpenWebSearch.EU, \url{https://doi.org/10.3030/101070014}).
}

\begin{raggedright}
  \small
  \bibliography{ecir25-rank-distillm-lit}
\end{raggedright}

\end{document}